# Paper self-citation:
# An unexplored phenomenon

Robin Haunschild[*] and Lutz Bornmann[*,**]

[*]*R.Haunschild@fkf.mpg.de, L.Bornmann@fkf.mpg.de*
0000-0001-7025-7256; 0000-0003-0810-7091
Information Retrieval Service CPT, Max Planck Institute for Solid State Research, Germany

[**]*bornmann@gv.mpg.de*
0000-0003-0810-7091
Science Policy and Strategy Department, Administrative Headquarters of the Max Planck Society, Germany

In this study, we investigated a phenomenon that one intuitively would assume does not exist: self-citations on the paper basis. Actually, papers citing themselves do exist in the Web of Science (WoS) database. In total, we obtained 44,857 papers that have self-citation relations in the WoS raw dataset. In part, they are database artefacts but in part they are due to papers citing themselves in the conclusion or appendix. We also found cases where paper self-citations occur due to publisher-made highlights promoting and citing the paper. We analyzed the self-citing papers according to selected metadata. We observed accumulations of the number of self-citing papers across publication years. We found a skewed distribution across countries, journals, authors, fields, and document types. Finally, we discuss the implications of paper self-citations for bibliometric indicators.

## 1. Introduction

In simple terms, a self-citation usually happens when the citing paper and the cited paper have at least one author in common (Carley, Porter, & Youtie, 2013). However, self-citations cannot only be studied on the basis of authors, but also on the level of journals, institutions, and countries (Bakare & Lewison, 2017). In these cases, self-citations are defined as citing papers and cited papers that have at least one journal, one institution, or one country in common. Most of the studies dealing with self-citations in previous bibliometric studies have considered self-citations on the author level. The accuracy of the common author identification at the citing and cited paper can be compromised by different authors having identical names (leading to overcounting self-citations, known as Type II errors) and by variations or misspellings in author names (leading to undercounting self-citations, known as Type I errors) (Evidence Ltd., 2007; Glänzel, Thijs, & Schlemmer, 2004).

Bibliometricians occasionally abstain from using author self-citations in citation analyses, arguing that author self-citations artificially boost citation counts and thereby unjustifiably enhance the authors' standing within the scientific community (Glänzel, Debackere, Thijs, & Schubert, 2006). However, there is no consensus in the bibliometric literature regarding the inclusion and exclusion of author self-citations in bibliometric studies. In citation analyses at low aggregation levels, such as those focusing on individual researchers, it might be appropriate to exclude self-citations (and to focus on the citation impact generated by papers of other authors). In studies at high aggregation levels such as institutions or countries, the impact of self-citations on the results may be minimal, indicating that exclusion may not be necessary. In the decision to include or exclude self-citations in citation analyses, one should consider that self-citations are widely accepted in some disciplines but viewed negatively in others. The choice to include or exclude them will have varying effects across different fields, "choosing to exclude them will affect some subject areas more than others" (Wilsdon et al., 2015, p. 32). Nosek et al. (2010) argue that only gratuitous self-citations, but not self-citations necessary for the specific content of a manuscript should be identified and corrected, but that the differentiation between both types in citation indexes is impossible.

One level of self-citations has not yet been studied in the bibliometric literature to the best of our knowledge: self-citations on the basis of individual papers. Although one may assume that these self-citations do not exist or do not make sense, we observed in the Web of Science (WoS, Clarivate) that they actually do exist. Paper self-citations happen when the citing paper and the cited paper are the same paper. With this study, we would like to draw attention to the phenomenon of paper self-citations by exploring its occurrence on the country, journal, field, and author level.

## 2. Dataset and methods

We used an April 2024 snapshot of the WoS that includes the Science Citation Index - Expanded (SCI-E), the Social Sciences Citation Index (SSCI), the Conference Proceedings Citation Index - Science (CPCI-S), the Conference Proceedings Citation Index - Social Science & Humanities (CPCI-SSH), and the Arts and Humanities Citation Index (AHCI) since 1980. The snapshot is licensed through, and made available by, the German Kompetenznetzwerk Bibliometrie (KB, Schmidt et al., 2024). This WoS raw dataset is stored in our database ivsdb at the Max Planck Computing and Data Facility (MPCDF).

All references cited in the WoS papers have a unique ID (UT accession number). For identifying paper self-citations, we selected those IDs where the IDs of the citing paper and the cited paper are the same. In total, we obtained 44,857 papers that have self-citation relations in the WoS raw dataset. In order to check whether these self-citations are merely database artefacts or if they really do exist in the scientific literature, we verified the self-citation relations in the WoS web-interface and checked them in the publisher's PDFs. In this study, we analyzed the set of 44,857 papers according to selected metadata.

## 3. Results

We checked examples of self-citing papers to obtain an insight into the reason for these self-citations:

- We found a self-citing paper that has been cited in the paper's appendix (Bendiak et al., 1997).
- We found a paper that has been cited in its own conclusions section (Hicks, 1997).
- We found a paper that has been cited by the journal's 'Frontispiece' (Chen et al., 2021; Huang, Wang, Tsai, Lin, & Lai, 2017) or 'Highlight' (Stalke, 1994).
- We found several papers that seem to be database errors (Carlomagno, Sánchez, Blommers, & Griesinger, 2003; Gevorgyan et al., 2016; Leijtens et al., 2015; Wang et al., 2021; Xia et al., 2021).

These examples show that self-citing papers seem to be a mixture of a real phenomenon one wouldn't expect and database errors. In order to have insights in the distribution of paper self-citations in science, we analyzed self-citing papers according to publication years, countries, journals, authors, WoS Subject Categories, and document types to obtain a deeper insight into this phenomenon. We used all units of analysis (e.g., countries, journals, author names, etc.) as they are provided in the WoS raw data by the KB.

*3.1 Analysis by publication years*

Figure 1 shows the number of self-citing papers in comparison with the number of all papers in the WoS (divided by 300) over the years. An accumulation of self-citing papers can be seen in the publication years 2006, 2017, and 2021 without particular accumulation of papers in the overall WoS. The strong decrease from 2023 to 2024 is because our data snapshot is from April

2024. The publication year 2023 is nearly but not completely full. Thus, it is unclear if the decrease in the number of self-citing papers from 2022 to 2023 is also an artefact or not.
Figure 2 shows the percentage of self-citing papers over the years. The main accumulations of self-citing papers also show up here in the publication years 2006, 2017, and 2021. In addition, minor accumulations are found for example in the publication years 1992 and 2010.

Figure 1. Number of self-citing papers in comparison with the number of all papers in the WoS (divided by 300) over the years

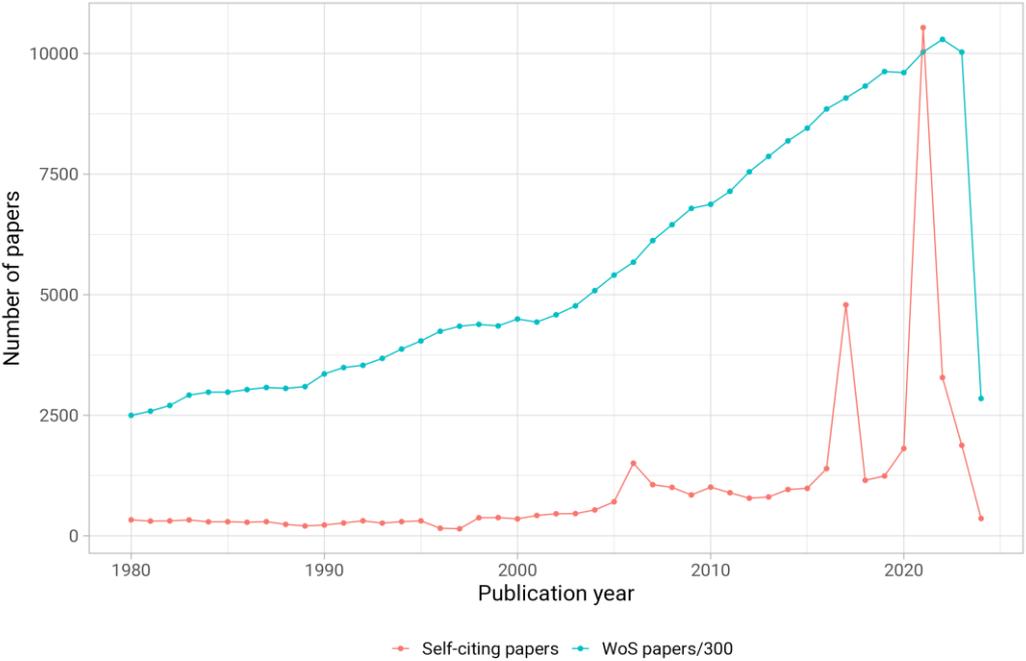

Figure 2. Percentage of self-citing papers over the years

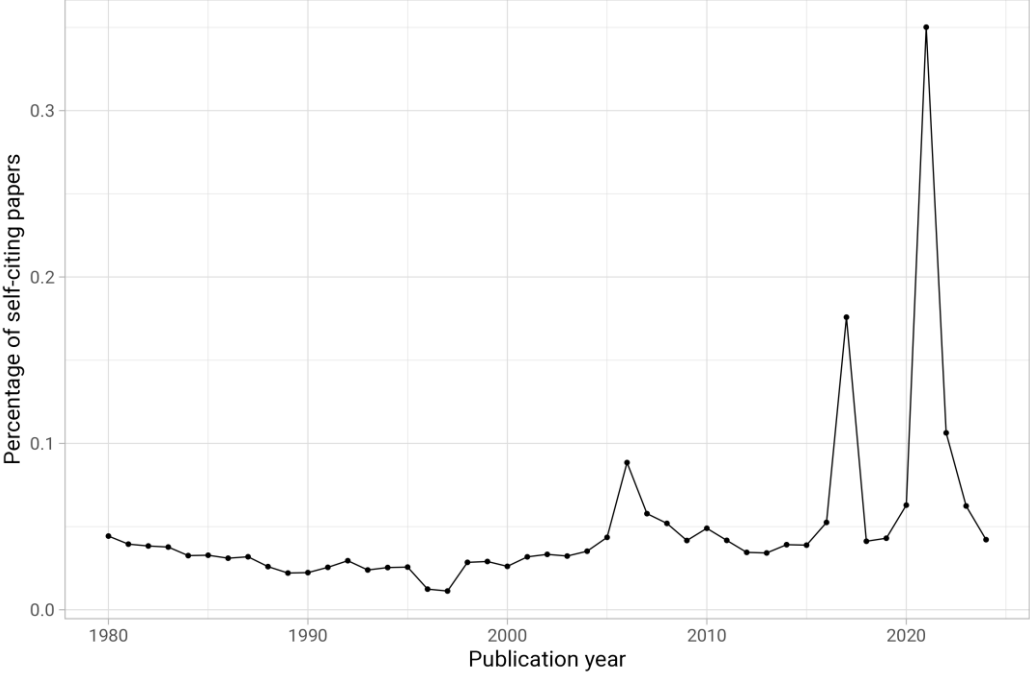

*3.2 Analysis by countries*

Table 1 shows the top 20 countries by number (upper half) and percentage (lower half) of self-citing papers. High-output countries like the United States, China, United Kingdom, and Germany head the list ordered by the number of self-citing papers whereas very low-output countries like the Falkland Islands, Equatorial Guinea, and Sint Maarten head the list ordered by the percentage of self-citing papers. The country with the highest overall output and number of self-citing papers in the top 20 list ordered by the percentage of self-citing papers is Panama with 29 self-citing papers. The only country with a percentage of self-citing papers higher than 1% is the Falkland Islands.

Table 1. Top 20 countries by number (upper half) and percentage (lower half) of self-citing papers

| Country | Number of self-citing papers | Number of papers | Percentage of self-citing papers |
|---|---|---|---|
| | ordered by number of self-citing papers | | |
| United States | 13,128 | 21,408,716 | 0.06 |
| China | 5,363 | 8,368,336 | 0.06 |
| United Kingdom | 4,493 | 5,668,517 | 0.08 |
| Germany | 3,601 | 4,628,285 | 0.08 |
| Canada | 2,296 | 2,943,905 | 0.08 |
| France | 2,135 | 3,154,701 | 0.07 |
| Japan | 2,039 | 3,870,548 | 0.05 |
| Australia | 1,970 | 2,107,785 | 0.09 |
| Italy | 1,927 | 2,636,702 | 0.07 |
| India | 1,506 | 2,155,528 | 0.07 |
| Spain | 1,488 | 1,973,751 | 0.08 |
| Netherlands | 1,446 | 1,529,741 | 0.09 |
| Switzerland | 1,098 | 1,130,827 | 0.10 |
| Sweden | 908 | 1,041,630 | 0.09 |
| Brazil | 901 | 1,234,385 | 0.07 |
| Korea, Republic of | 845 | 1,527,237 | 0.06 |
| Poland | 731 | 905,680 | 0.08 |
| Russian Federation | 718 | 1,295,418 | 0.06 |
| Denmark | 676 | 639,195 | 0.11 |
| Turkey | 654 | 831,664 | 0.08 |
| | ordered by percentage of self-citing papers | | |
| Falkland Islands | 3 | 272 | 1.10 |
| Equatorial Guinea | 2 | 222 | 0.90 |
| Sint Maarten | 1 | 114 | 0.88 |
| Sao Tome and Principe | 1 | 147 | 0.68 |
| Cayman Islands | 1 | 257 | 0.39 |
| Maldives | 2 | 557 | 0.36 |
| Timor-Leste/Portuguese Timor | 1 | 333 | 0.30 |
| Djibouti | 1 | 362 | 0.28 |
| Palau | 1 | 362 | 0.28 |
| Vanuatu | 2 | 754 | 0.27 |

| Micronesia, Federated States of | 1 | 378 | 0.26 |
| Vatican City State | 2 | 808 | 0.25 |
| Bermuda | 3 | 1,229 | 0.24 |
| Panama | 29 | 12,118 | 0.24 |
| Liberia | 3 | 1,293 | 0.23 |
| Dominica | 1 | 460 | 0.22 |
| Mauritania | 2 | 997 | 0.20 |
| Montenegro | 10 | 6,223 | 0.16 |
| El Salvador | 3 | 1,950 | 0.15 |
| Fiji | 8 | 5,400 | 0.15 |

*3.3 Analysis by journals*

Figure 3 shows a log-log plot of the number of journals and the number of self-citing papers. In total, we found 7,943 journals with at least one self-citing paper. The distribution is skewed: Only a few journals have many self-citing papers. It takes only eight journals (0.1%) to cover the top 10%, 129 journals (1.6%) to cover the top third, 365 journals (4.6%) to cover the first half, and 932 journals (11.7%) to cover the first two thirds of all 7,943 journals with at least one self-citing paper.

Figure 3. Log-log plot of the number of journals and the number of self-citing papers

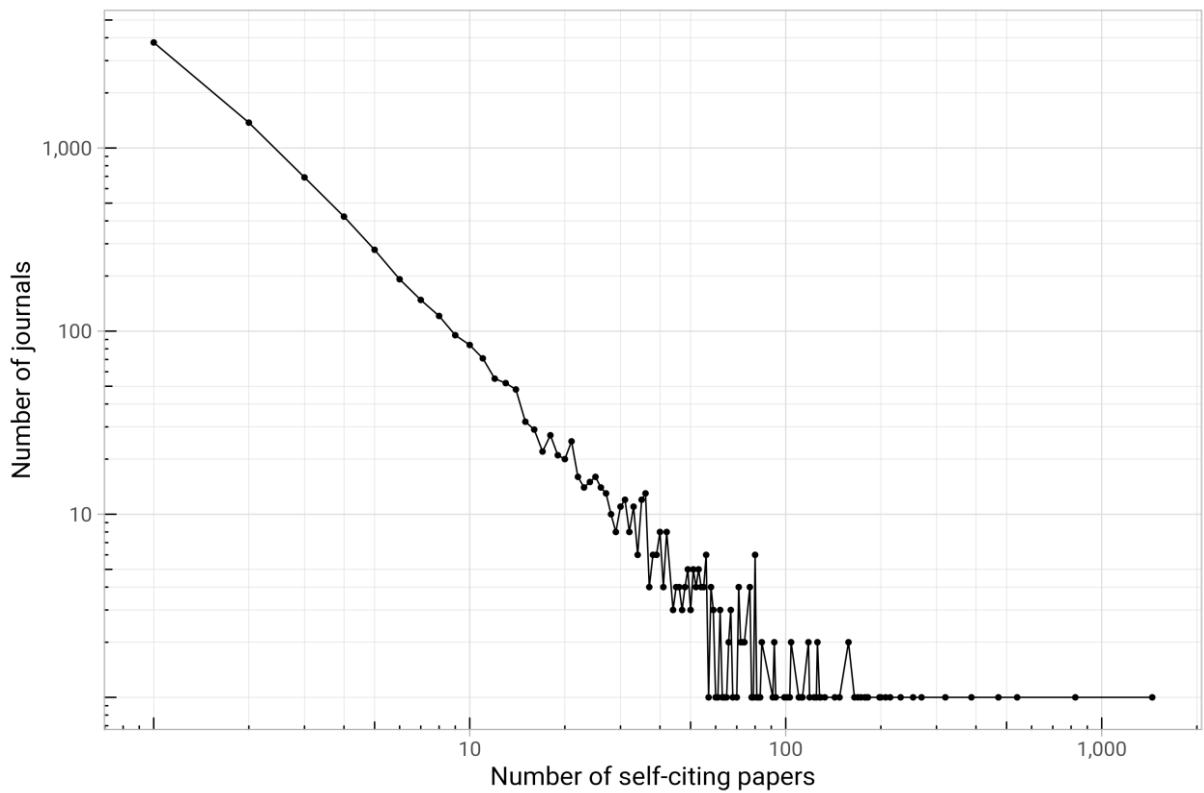

Table 2 shows the journals with at least 100 self-citing papers ordered by the percentage of self-citing papers. Most self-citing papers were published in *Annals of Thoracic Surgery*, *ChemistrySelect*, and *Cochrane Database of Systematic Reviews* whereas the highest percentages of self-citing papers with at least 100 self-citing papers can be found in *Quantitative Economics*, *Journal of African Media Studies*, and *Journal of Screenwriting*.

Table 2. Journals with at least 100 self-citing papers ordered by the percentage of self-citing papers

| Journal | Number of self-citing papers | Number of papers | Percentage of self-citing papers |
| --- | --- | --- | --- |
| *Quantitative Economics* | 165 | 416 | 39.66 |
| *Journal of African Media Studies* | 158 | 417 | 37.89 |
| *Journal of Screenwriting* | 118 | 463 | 25.49 |
| *Revista Colombiana de Entomologia* | 126 | 713 | 17.67 |
| *I-Perception* | 207 | 2,046 | 10.12 |
| *Prehospital and Disaster Medicine* | 129 | 1,771 | 7.28 |
| *Neurocirugia* | 110 | 1,620 | 6.79 |
| *Livestock Production Science* | 169 | 2,655 | 6.37 |
| *ChemistrySelect* | 825 | 15,278 | 5.40 |
| *Cochrane Database of Systematic Reviews* | 540 | 13,411 | 4.03 |
| *Annals of Thoracic Surgery* | 1,445 | 39,136 | 3.69 |
| *Kardiologia Polska* | 173 | 4,773 | 3.62 |
| *Chemelectrochem* | 148 | 4,566 | 3.24 |
| *Econometrica* | 101 | 3,118 | 3.24 |
| *Behavioral Ecology* | 125 | 4,715 | 2.65 |
| *Journal of Food Processing and Preservation* | 143 | 6,764 | 2.11 |
| *American Economic Review* | 158 | 8,272 | 1.91 |
| *Advanced Energy Materials* | 128 | 7,417 | 1.73 |
| *European Urology* | 269 | 18,066 | 1.49 |
| *Advanced Science* | 113 | 7,922 | 1.43 |
| *Advanced Functional Materials* | 253 | 22,625 | 1.12 |
| *Water Resources Research* | 198 | 17,915 | 1.11 |
| *Crop Science* | 179 | 16,215 | 1.10 |
| *Journal of Geophysical Research-Oceans* | 178 | 16,155 | 1.10 |
| *Ecology and Evolution* | 119 | 10,830 | 1.10 |
| *American Journal of Medical Genetics Part A* | 104 | 11,618 | 0.90 |
| *Journal of Geophysical Research-Atmospheres* | 214 | 27,269 | 0.78 |
| *Geophysical Research Letters* | 320 | 45,576 | 0.70 |
| *Small* | 123 | 17,585 | 0.70 |
| *European Journal of Organic Chemistry* | 126 | 18,092 | 0.70 |
| *Advanced Materials* | 118 | 27,569 | 0.43 |
| *Angewandte Chemie-International Edition* | 231 | 58,118 | 0.40 |
| *Journal of Applied Polymer Science* | 200 | 52,618 | 0.38 |
| *Chemistry-A European Journal* | 133 | 38,707 | 0.34 |
| *Physical Review Letters* | 387 | 116,784 | 0.33 |
| *Biochemistry* | 182 | 58,775 | 0.31 |
| *Physical Review B* | 471 | 199,353 | 0.24 |
| *Astronomy & Astrophysics* | 112 | 68,026 | 0.16 |
| *Physical Review E* | 103 | 67,425 | 0.15 |
| *Scientific Reports* | 104 | 203,117 | 0.05 |

*3.4 Analysis by authors*

Figure 4 shows the distribution of the number of authors versus the number of self-citing papers. Five author names are connected to the highest number of self-citing papers (n=88). As we have no good author-name-disambiguation algorithm available in WoS, we can't be sure if these names correspond to a single person each or multiple persons. It might also be the case that some author's synonyms are not captured so that the analysis of author names can only be seen as a rough estimate.

Figure 4. Distribution of the number of authors versus the number of self-citing papers

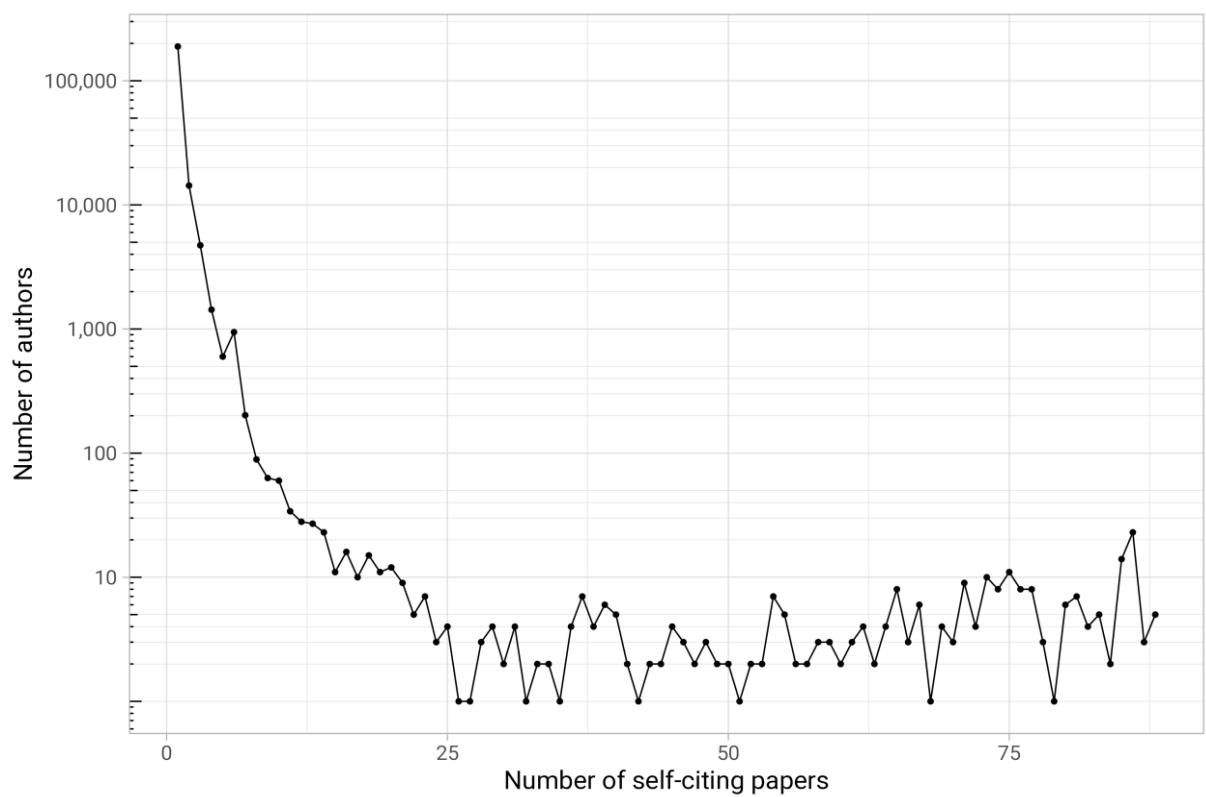

*3.5 Analysis by WoS Subject Categories*

Table 3 shows the 20 WoS Subject Categories with the most self-citing papers ordered by the number (upper half) and percentage (lower half) of self-citing papers. The WoS Subject Categories with the most self-citing papers are from different fields within chemistry, physics, medicine, and biology. The list of the WoS Subject categories with the highest percentages of self-citing papers contains many more fields from social sciences (e.g., Cultural Studies and Social Work) besides biomedical fields (e.g., Respiratory System and Evolutionary Biology). The results of Glänzel, et al. (2004) for self-citations on the author level show that "Biomedical research, Chemistry, and Mathematics have the highest share of self-citations (60%) in the year of publication. The field 'Social sciences I' has by far the lowest one (32.5%)" (Glänzel, et al., 2004, p. 69). Snyder and Bonzi (1998) found that the number of author self-citations is greater in the physical science (15%) than in the social sciences (6%) or the humanities (6%). At least in part, these author self-citations can now be attributed to paper self-citations.

Table 3. 20 WoS Subject Categories with the most self-citing papers ordered by the number (upper half) and percentage (lower half) of self-citing papers

| WoS Subject Category | Number of self-citing papers | Number of WoS papers | Percentage of self-citing papers |
|---|---|---|---|
| ordered by the number of self-citing papers | | | |
| Chemistry, Multidisciplinary | 2,821 | 2,587,190 | 0.11 |
| Materials Science, Multidisciplinary | 2,527 | 3,117,659 | 0.08 |
| Surgery | 2,297 | 1,693,860 | 0.14 |
| Cardiac & Cardiovascular Systems | 2,175 | 1,332,432 | 0.16 |
| Physics, Applied | 1,800 | 2,402,559 | 0.07 |
| Respiratory System | 1,734 | 647,728 | 0.27 |
| Chemistry, Physical | 1,510 | 1,875,520 | 0.08 |
| Biochemistry & Molecular Biology | 1,494 | 2,732,834 | 0.05 |
| Physics, Condensed Matter | 1,429 | 1,232,159 | 0.12 |
| Engineering, Electrical & Electronic | 1,392 | 4,220,510 | 0.03 |
| Medicine, General & Internal | 1,334 | 1,841,512 | 0.07 |
| Economics | 1,172 | 863,849 | 0.14 |
| Environmental Sciences | 1,137 | 1,652,488 | 0.07 |
| Ecology | 1,067 | 608,778 | 0.18 |
| Geosciences, Multidisciplinary | 1,010 | 806,559 | 0.13 |
| Nanoscience & Nanotechnology | 985 | 815,037 | 0.12 |
| Neurosciences | 935 | 1,781,375 | 0.05 |
| Pharmacology & Pharmacy | 901 | 1,742,059 | 0.05 |
| Cell Biology | 814 | 1,446,761 | 0.06 |
| Computer Science, Theory & Methods | 753 | 1,318,424 | 0.06 |
| ordered by the percentage of self-citing papers | | | |
| Cultural Studies | 164 | 46,390 | 0.35 |
| Limnology | 234 | 70,224 | 0.33 |
| Film, Radio, Television | 427 | 155,843 | 0.27 |
| Respiratory System | 1,734 | 647,728 | 0.27 |
| Medical Ethics | 75 | 34,832 | 0.22 |
| Evolutionary Biology | 440 | 212,928 | 0.21 |
| Ecology | 1,067 | 608,778 | 0.18 |
| Ornithology | 88 | 51,333 | 0.17 |
| Social Sciences, Mathematical Methods | 152 | 89,916 | 0.17 |
| Andrology | 45 | 26,735 | 0.17 |
| Social Work | 170 | 101,206 | 0.17 |
| Cardiac & Cardiovascular Systems | 2,175 | 1,332,432 | 0.16 |
| Emergency Medicine | 212 | 135,188 | 0.16 |
| Meteorology & Atmospheric Sciences | 631 | 411,557 | 0.15 |
| Logic | 50 | 34,749 | 0.14 |
| Industrial Relations & Labor | 91 | 64,962 | 0.14 |

| Mathematics, Interdisciplinary Applications | 406 | 290,819 | 0.14 |
| Psychology, Experimental | 401 | 287,571 | 0.14 |
| Communication | 201 | 145,192 | 0.14 |
| Economics | 1,172 | 863,849 | 0.14 |

*3.6 Analysis by document types*

Table 4 shows the distribution of self-citing papers across the WoS document types (only the primary document type is considered) ordered by the number of self-citing papers. Although the percentage is very low, the highest number of self-citing papers is found for the most frequently occurring document type 'Article'. The highest percentages are found for the seldom occurring document types 'Item Withdrawal' and 'Retraction'. In those cases, the retraction notices might be confused with the retracted papers. Similar reasons can be expected for the document types 'Correction', 'Correction, Addition', 'Reprint', and 'Expression of Concern'.

Table 4. Distribution of self-citing papers across the WoS document types (only the primary document type is considered) ordered by the number of self-citing papers

| Document type | Number of self-citing papers | Number of WoS papers | Percentage of self-citing papers |
| --- | --- | --- | --- |
| Article | 31,201 | 44,802,527 | 0.07 |
| Correction | 6,159 | 395,752 | 1.56 |
| Review | 2,496 | 2,522,206 | 0.10 |
| Proceedings Paper | 1,428 | 7,632,141 | 0.02 |
| Editorial Material | 1,329 | 3,082,569 | 0.04 |
| Letter | 962 | 1,905,649 | 0.05 |
| Retraction | 556 | 17,392 | 3.20 |
| Note | 136 | 840,918 | 0.02 |
| Item Withdrawal | 129 | 391 | 32.99 |
| Book Review | 110 | 3,312,905 | 0.00 |
| Correction, Addition | 99 | 88,441 | 0.11 |
| Reprint | 87 | 17,870 | 0.49 |
| Meeting Abstract | 72 | 8,114,286 | 0.00 |
| News Item | 44 | 591,925 | 0.01 |
| Biographical-Item | 27 | 141,501 | 0.02 |
| Discussion | 10 | 39,146 | 0.03 |
| Expression of Concern | 10 | 1,322 | 0.76 |
| Database Review | 1 | 1,448 | 0.07 |
| Fiction, Creative Prose | 1 | 45,175 | 0.00 |

## 4. Discussion and conclusions

This study dealt with a phenomenon in bibliometrics that has not been studied up to now: self-citations on the basis of individual papers. Reasons for missing previous studies may lie (1) in the fact that this phenomenon should actually not exist: In the usual process of citing papers, one assumes that it does not make sense to cite the paper one currently writes (Tahamtan & Bornmann, 2018). We inspected examples in the WoS to explore possible reasons for papers self-citations and found that self-citing papers may be a mixture of a real phenomenon (e.g., the supporting information cites the main manuscript) and database errors. (2) Another reason may be that the phenomenon is a rare event: We found in most publication years percentages of less than 0.1% of self-citing papers. We observed these low percentages and numbers not only for publication years, but also for countries, journals, authors, subject categories, and document types. Studies that have analyzed self-citations on the author-level report significantly higher numbers, e.g., that around 10% of cited references are self-citations (Garfield, 1979; King, Bergstrom, Correll, Jacquet, & West, 2017). Aksnes (2003) investigated the role of author self-citation in the scientific production of Norway. He found that 36% of all citations represent self-citations.

For Glänzel, et al. (2006), there is no justification for universally condemning author self-citations or excluding them from macro statistics. Instead, bibliometric indicators based on self-citations may be useful for understanding communication patterns. Self-citations typically indicate the use of an author's previous results in a new publication, shorten publications by referencing previously published methodologies, or make author's background materials in "grey" literature more visible. Similar arguments can be found in Pendlebury (2008). Davis (2009) emphasizes that high levels of self-citations indicate a narrow specialty "rather than a willful attempt to work the system" (p. 8). However, if self-citations are shown to be (completely) independent of the number of external citations, citation indicators should indeed be adjusted to exclude self-citations, as they could be potential means of manipulation.

We consider the assessments of author self-citations by Glänzel, et al. (2006) and Pendlebury (2008) to be appropriate for paper self-citations as well. In most citation analyses, there is no necessity to exclude paper self-citations in bibliometric studies. To the best of our knowledge, there is only one context in bibliometrics, where it is necessary to exclude these citations: in the calculation of the disruption index (Leibel & Bornmann, 2024) and similar ones. This index has been recently proposed to measure the disruptiveness of research published in a paper based on its network of citing and cited papers (Wu, Wang, & Evans, 2019). The disruption index is a value between -1 and 1 without paper self-citations with 1 being the most disruptive value and -1 the most consolidating value. However, when paper self-citations are included, the disruption index can (erroneously) assume values between -2 and 2.

**Open science practices**
As WoS data are proprietary, we are unfortunately not allowed to share the data.

**Acknowledgments**
The ivsdb data stem from a bibliometrics database based on data provided by the German Kompetenznetzwerk Bibliometrie (KB, Competence Network Bibliometrics, funded by BMBF via grant 01PQ17001, see: http://www.bibliometrie.info/). The data are derived from the Science Citation Index - Expanded (SCI-E), the Social Sciences Citation Index (SSCI), the Conference Proceedings Citation Index - Science (CPCI-S), the Conference Proceedings Citation Index - Social Science & Humanities (CPCI-SSH), and the Arts and Humanities Citation Index (AHCI), provided by Clarivate and updated in calendar week 17 of 2024.

**Author contributions**
Conceptualization: RH & LB
Data curation: RH
Investigation: RH
Methodology: RH & LB
Visualization: RH
Writing – original draft: RH & LB

Writing – review & editing: RH & LB

**Competing interests**
The authors declare that they have no competing interests.

**Funding information**
Access to bibliometric data was enabled via grant 01PQ17001.